\documentclass[12pt]{article}
\usepackage{graphicx}
\begin{document}


\begin{flushright}
from http://arxiv.org/abs/1203.4539
\end{flushright}

\vspace{20mm}

\begin{center}
{\Large \bf Poincar\'e Sphere and Decoherence Problems}

\vspace{3ex}

Y. S. Kim \footnote{electronic address: yskim@umd.edu}\\
Center for Fundamental Physics, University of Maryland,\\
College Park, Maryland 20742, U.S.A.

\end{center}

\vspace{3ex}
\begin{abstract}
Henri Poincar\'e formulated the mathematics of the Lorentz transformations,
known as the Poincar\'e group.  He also formulated the Poincar\'e sphere
for polarization optics.  It is shown that these two mathematical
instruments can be combined into one mathematical device which can address
the internal space-time symmetries of elementary particles, decoherence
problems in polarization optics, entropy problems, and Feynman's rest of
the universe.

\end{abstract}
\vspace{30mm}

\noindent Presented at the Fedorov Memorial Symposium:
International Conference "Spins and Photonic Beams at Interface,"
dedicated to the 100th anniversary of F.I.Fedorov (1911-1994)
(Minsk, Belarus, 2011)

\newpage

\section{Introduction}\label{intro}
It was Henri Poincar\'s who worked out the mathematics of Lorentz
transformations before Einstein and Minkowski.  Poincar\'e's interest
in mathematics covers many other areas of physics.
\par
Among them, his Poincar\'e sphere serves as geometrical representation
of the Stokes parameters.  While there are four Stokes parameters, the
traditional Poincar\'e sphere is based on only three of those
parameters~\cite{azzam77,born80,bros98}.
\par
In this report, we show that the concept of the Poincar\'e sphere can be
extended to accommodate all four Stokes parameters, and show that this
extended Poincar\'e sphere can be used as a representation of the Lorentz
group applicable to the four-component Minkowski space.  In this way,
it is possible to use the Poincar\'s sphere as a picture of the internal
space-time symmetries as defined by Wigner in 1939~\cite{wig39}.
\par

\begin{figure}[thb]
\centerline{\includegraphics[scale=1.6]{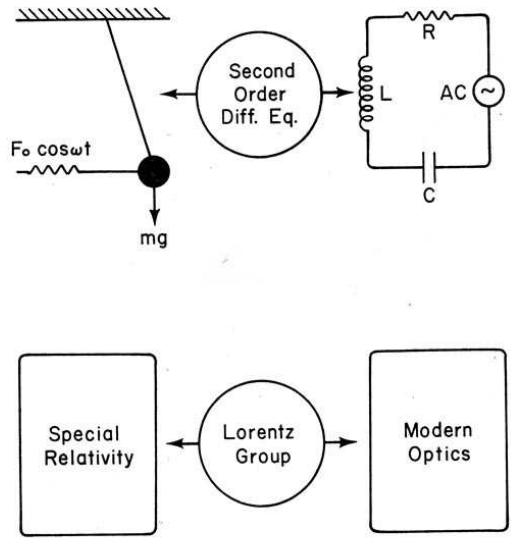}}
\caption{One mathematical instrument for two different branches of physics.
As one second-order differential equation takes care of the resonances in
the oscillator system and also in the $LCR$ circuit, the Lorentz group can
serve as the basic mathematical language for special relativity and modern
optics.}\label{resonance}
\end{figure}

Throughout the paper, we shall use the two-by-two matrix formulation of the
Lorentz group~\cite{naimark54,naimark64}.  The basic advantage is that this
representation is the natural language for the two-by-two representation of
the four Stokes parameters.  Indeed, this two-by-two representation speaks
one language applicable to two different branches of physics, as is
illustrated in Fig.~\ref{resonance}.
\par

Let us consider a particle moving along a given direction.  We can rotate
the system around this direction, and boost along this direction.  We can
also rotate around and boost along two orthogonal directions perpendicular
to the direction of the momentum.  These operations can be written in the
form of two-by-two matrices.  In this way, we can construct a two-by-two
representation of the Lorentz group.

\par
In this report, we use the same set of two-by-two matrices to perform
transformations on the Stokes parameters and the Poincar\'e sphere.  While the
original three-dimensional Poincar\'e sphere is based  only on  three of the
four Stokes parameters, it is possible to extend its geometry to accommodate
all four parameters.  With these parameters, we can address the degree of
polarization between the two orthogonal components of electric field
perpendicular to the momentum.   In order to deal with this problem, we
introduce the decoherence parameter derivable from the traditional degree of
polarization.  Since this parameter is greater than zero and smaller than
one, we can write it as $\sin\chi$, and define $\chi$ as the decoherence
angle.

\par
One remarkable aspect of this decoherence parameter is that it is invariant
under Lorentz transformations.  If the symmetry of the Poincar\'e sphere is
translated into the space-time symmetry, the decoherence parameter remains
Lorentz-invariant like the particle mass in the space-time symmetry.
\par
Furthermore,  to every $\sin\chi$, there is $\cos\chi$, and their symmetry
is well known.  Thus we can consider another Poincar\'e sphere with the same
decoherence angle.  The second Poincar\'e sphere could serve as another
illustrative example of Feynman's rest of the universe~\cite{fey72,hkn99ajp}.

\par
In Sec.~\ref{pew}, it was noted that Poincar\'e was the first one to
formulate group of Lorentz transformations for the space-time variables.
This transformation law can be translated into the language of Naimark's
two-by-two Naimark representation.  Einstein later derived his $E = mc^2$
by advancing the concept of the energy-momentum four-vector obeying the
same transformation law as that of the space-time coordinates.  Thus,
the two-by-two representation is possible also for the momentum-energy
four-vector.
\par
When Einstein was formulating his special relativity, he did not consider
internal space-time structures or symmetries of the particles.  Also in
Sec.~\ref{pew}, it is shown possible to study Wigner's little groups using
the two-by-two representation.  Wigner's little groups dictate the internal
space-time symmetries of elementary particles.  They are the subgroups of
the Lorentz group whose transformations leave the four-momentum of the
given particle invariant~\cite{wig39,knp86}.
\par
In Sec.~\ref{jones}, we first note that the same two-by-two matrices are
applicable to two-component Jones vectors. However, the Jones vector cannot
address the problem of coherence between two transverse components of the
optical beam.  This is why the four-component Stokes parameters are needed.
The same Lorentz group can be used for the four-component Stokes parameters,
and Stokes parameters are like four-vectors, as in the case of energy-momentum
four vectors.  It is noted however, the degree of coherence remains invariant
under Lorentz transformations.
\par
In Sec.~\ref{poincs}, we use the geometry of the Poincar\'e sphere to describe
the Stokes parameters and their transformation properties.  It is shown that
three different radii are needed to describe them fully.   The traditional
radius lies between the maximum and minimum radii.  It is shown that
$\left(S_0^2 - R^2\right)$ remains lorentz-invariant, where $S_0$ and $R$
are the maximum radius and the traditional radius respectively.  This
Lorentz-invariant quantity is dictated by the decoherence parameter.  The entropy
question is also discussed, and this entropy is shown to be Lorentz-invariant.
\par
In Sec.~\ref{restof}, it is shown possible to define another Poincar\'e
sphere in order to deal with variations of the decoherence parameter and
the entropy.  It is shown that this second Poincar\'e sphere can serve
as another example of Feynman's rest of the universe~\cite{fey72}.

\par

In the the Appendix, we shoe how the two-by-two transformation matrix can
be translated into the four-by-four matrix applicable to the Missourian
four-vector.

\section{Poincar\'e Group, Einstein, and Wigner}\label{pew}
The Lorentz group starts with a group of four-by-four matrices
performing Lorentz transformations on the Minkowskian vector
space of $(t, z, x, y),$ leaving the quantity
\begin{equation}\label{4vec02}
t^2 - z^2 - x^2 - y^2
\end{equation}
invariant.  It is possible to perform this transformation using two-by-two
representations~\cite{naimark54}.  This mathematical aspect is known as
the $SL(2,c)$ as the universal covering group for the Lorentz group.

\par
In this two-by-two representation, we write the four-vector as a matrix
\begin{equation}
X = \pmatrix{t + z  &  x - iy \cr x + iy & t - z} .
\end{equation}
Then its determinant is precisely the quantity given in Eq.(\ref{4vec02}).
Thus the Lorentz transformation on this matrix is a determinant-preserving
transformation.  Let us consider the transformation matrix as
\begin{equation}\label{g22}
 G = \pmatrix{\alpha & \beta \cr \gamma & \delta}, \qquad G^{\dagger} =
  \pmatrix{\alpha^* & \gamma^* \cr \beta^* & \delta^*} ,
\end{equation}
with
\begin{equation}
    \det{(G)} = 1.
\end{equation}
This matrix has six independent parameters.   The group of these $G$ matrices
is known to be locally isomorphic to the group of four-by-four matrices
performing Lorentz transformations on the four-vector $(t, z, x, y)$.  In
other word, for each $G$ matrix there is a corresponding four-by-four
Lorentz-transform matrix, as is illustrated in the Appendix.
\par

The matrix $G$ is not a unitary matrix, because its Hermitian conjugate is
not always its inverse.   The group can have a unitary subgroup called $SU(2)$
performing rotations on electron spins.  As far as we can see, this
$G$-matrix formalism was first presented by Naimark in 1954~\cite{naimark54}.
Thus, we call this formalism the Naimark representation of the Lorentz group.
We shall see first that this representation is convenient for studying
space-time symmetries of particles.  We shall then note that this Naimark
representation is the natural language for the Stokes parameters in
polarization optics.

\par
With this point in mind, we can now consider the transformation
\begin{equation}\label{naim}
X' = G X G^{\dagger}
\end{equation}
Since $G$ is not a unitary matrix, it is not a unitary transformation.
In order to tell this difference, we call this ``Naimark transformation.''
This expression can be written explicitly as
\begin{equation}\label{lt01}
\pmatrix{t' + z' & x' - iy' \cr x + iy & t' - z'}
 = \pmatrix{\alpha & \beta \cr \gamma & \delta}
  \pmatrix{t + z & x - iy \cr x + iy & t - z}
  \pmatrix{\alpha^* & \gamma^* \cr \beta^* & \delta^*} ,
\end{equation}
\par
For this transformation, we have to deal with four complex numbers.  However,
for all practical purposes, we may work with two Hermitian matrices
\begin{equation}\label{herm11}
Z(\delta) = \pmatrix{e^{i\delta/2} & 0 \cr 0 & e^{-i\delta/2}}, \qquad
R(\delta) = \pmatrix{\cos(\theta/2)  & -\sin(\theta/2) \cr
     \sin(\theta/2) & \cos(\theta/2)} ,
\end{equation}
and two symmetric matrices
\begin{equation}\label{symm11}
B(\mu) = \pmatrix{e^{\mu/2} & 0 \cr 0 & e^{-\mu/2}}, \qquad
S(\lambda) = \pmatrix{\cosh(\lambda/2)  & \sinh(\lambda/2) \cr
     \sinh(\lambda/2) & \cosh(\lambda/2)}
\end{equation}
The two Hermitian matrices in Eq.(\ref{herm11}) lead to rotations around
the $z$ and $y$ axes respectively.  The symmetric matrices in Eq.(\ref{symm11})
perform Lorentz boosts along the $z$ and $x$ directions.

\par

Repeated applications of these four matrices will lead to the most general
form the of the most general form of the $G$ matrix of Eq.(\ref{g22}) with
six independent parameters.  For each two-by-two Naimark transformation,
there is a four-by-four matrix performing the corresponding Lorentz
transformation on the four-component four vector.  In the appendix, the
four-by-four equivalents are given for the matrices of Eq.(\ref{herm11})
and Eq.(\ref{symm11}).

\par

It was Einstein who defined the energy-momentum four vector, and showed
that it also has the same Lorentz-transformation law as the space-time
four-vector.  We write the energy-momentum four vector as
\begin{equation}\label{momen11}
P = \pmatrix{E + p_z & p_x - ip_y \cr p_x + ip_y & E - p_z} ,
\end{equation}
with
\begin{equation}
\det{(P)} = E^2 - p_x^2 - p_y^2 - p_z^2,
\end{equation}
which means
\begin{equation}\label{mass}
\det{{p}} = m^2,
\end{equation}
where $m$ is the particle mass.
\par
Now Einstein's transformation law can be written as
 \begin{equation}
 P' = G M G^{\dagger} ,
 \end{equation}
or explicitly
\begin{equation}\label{lt03}
\pmatrix{E' + p_z' & p_x' - ip_y' \cr p'_x + ip'_y & E' - p'_z}
 = \pmatrix{\alpha & \beta \cr \gamma & \delta}
  \pmatrix{E + p_z & p_x - ip_y \cr p_x + ip_y & E - p_z}
  \pmatrix{\alpha^* & \gamma^* \cr \beta^* & \delta^*} ,
\end{equation}

\par
Later in 1939~\cite{wig39}, Wigner was interested in constructing subgroups
of the Lorentz group whose transformations leave a given four-momentum
invariant, and called these subsets ``little groups.'' Thus, Wigner's
little group consists of two-by-two matrices satisfying
\begin{equation}\label{wigcon}
P = W P W^{\dagger} .
\end{equation}
This two-by-two $W$ matrix is not an identity matrix, but tells about
internal space-time symmetry of the particle with a given energy-momentum
four-vector.  This aspect was not known when Einstein formulated his special
relativity in 1905.

\par

If its determinant is a positive number, the $P$ matrix can be
brought to the form
\begin{equation}\label{massive}
         P = \pmatrix{1 & 0 \cr 0 & 1},
\end{equation}
corresponding to a massive particle at rest
\par
If the determinant is negative, it can be brought to the form
\begin{equation}\label{superlum}
         P = \pmatrix{1 & 0 \cr 0 & -1} ,
\end{equation}
corresponding to an imaginary-mass particle moving faster than light
along the $z$ direction, with its vanishing energy component.

\par
If the determinant if zero, we may write $P$ as
\begin{equation}\label{massless}
         P = \pmatrix{1 & 0 \cr 0 & 0} ,
\end{equation}
corresponding to a massless particle moving along the $z$ direction.
\par
For all three of the above cases, the matrix of the form
\begin{equation}\label{22rotz}
Z(\delta) = \pmatrix{e^{i\delta/2} & 0 \cr 0 & e^{-i\delta/2}}
\end{equation}
will satisfy the Wigner condition of Eq.(\ref{wigcon}).  This matrix
corresponds to rotations around the $z$ axis, as is shown in the Appendix.

\par
For the massive particle with the four-momentum of Eq.(\ref{massive}),
the Naimark transformations with the rotation matrix of the form
\begin{equation}
R(\theta) = \pmatrix{\cos(\theta/2) & -\sin(\theta/2) \cr
\sin(\theta/2) & \cos(\theta/2)} ,
\end{equation}
also leaves the $P$ matrix of Eq.(\ref{massive}) invariant.  Together with
the $Z(\delta)$ matrix, this rotation matrix lead to the subgroup consisting
of unitary subset of the $G$ matrices.  The unitary subset of $G$ is $SU(2)$
corresponding to the three-dimensional rotation group dictating the spin
of the particle~\cite{knp86}.
\par
For the massless case, the transformations with the triangular matrix of
the form
\begin{equation}
\pmatrix{1 & \gamma \cr 0 & 1}
\end{equation}
leaves the momentum matrix of Eq.(\ref{massless}) invariant.  The physics of
this matrix has a stormy history, and the variable $\gamma$ leads to gauge
transformation applicable to massless particles~\cite{hks82,kiwi90jm}.
\par
For a superluminal particle with it imaginary mass, the $W$ matrix of the form
\begin{equation}
S(\lambda) = \pmatrix{\cosh(\lambda/2) & \sinh(\lambda/2) \cr \sinh(\lambda/2)
 & \cosh(\lambda/2)}
\end{equation}
will leave the four-momentum of Eq.(\ref{superlum}) invariant.  This unobservable
particle does not appear to have observable internal space-time degrees of freedom.

\par

Table~\ref{tab11} summarizes the transformation matrices for Wigner's subgroups for
massive, massless, and superluminal transformations.  Of course, it is a challenging
problem to have one expression for all those three cases, and this problem has
been addressed in the literature~\cite{bk10jmo}.

\begin{table}[h]
\caption{Wigner's Little Groups.  The little groups are the subgroups of
the Lorentz group whose transformations leave the four-momentum of a given
particle invariant.  They thus define the internal space-time symmetries of
particles.  The four-momentum remains invariant under the rotation around it.
In addition, they remain invariant under the following transformations.  They
are different for massive, massless, and superluminal particles.}\label{tab11}
\vspace{2mm}

\begin{center}
\begin{tabular}{lcl}
\hline
\hline \\[0.5ex]
 Particle mass &  Four-momentum  &  Transform matrices \\[1.0ex]
\hline\\
massive  & $\pmatrix{1 & 0 \cr 0 & 1}$
&
$\pmatrix{\cos(\theta/2) & -\sin(\theta/2)\cr \sin(\theta/2) & \cos(\theta/2)}$
\\[4ex]
Massless  &
$\pmatrix{1 & 0 \cr 0 & 0}$
& $\pmatrix{1 & \gamma \cr 0 & 1}$
\\[4ex]
Imaginary mass &
$\pmatrix{1 & 0\cr 0 & -1}$
&  $\pmatrix{\cosh(\lambda/2) & \sinh(\lambda/2) \cr \sinh(\lambda/2) & \cosh(\lambda/2)}$
\\[4ex]
\hline
\hline\\[-0.8ex]
\end{tabular}
\end{center}
\end{table}

\section{Jones Vectors and Stokes Parameters}\label{jones}

In studying the polarized light propagating along the $z$ direction, the
traditional approach is to consider the $x$ and $y$ components of
the electric fields.  Their amplitude ratio and the phase difference
determine the state of polarization.  Thus, we can change the polarization
either by adjusting the amplitudes, by changing the relative phase,
or both.  For convenience, we call the optical device which changes
amplitudes an ``attenuator'' and the device which changes the relative
phase a ``phase shifter.''
\par
The traditional language for this two-component light is the Jones-matrix
formalism which is discussed in standard optics textbooks~\cite{saleh07}.
In this formalism, the above two components are combined into one column
matrix with the exponential form for the sinusoidal function
\begin{equation}\label{jvec11}
\pmatrix{\psi_1(z,t) \cr \psi_2(z,t)} =
\pmatrix{a \exp{\left\{i(kz - \omega t + \phi_{1})\right\}}  \cr
b \exp{\left\{i(kz - \omega t + \phi_{2})\right\}}} .
\end{equation}
This column matrix is called the Jones vector.
\par

When the beam goes through a medium with different values of indexes of
refraction for the $x$ and $y$ directions, we have to apply the matrix
\begin{equation}\label{phase3}
\pmatrix{e^{i\delta_{1}} & 0 \cr 0 & e^{i\delta_{2}}}
= e^{i(\delta_{1} + \delta_{2})/2}
\pmatrix{e^{-i\delta/2} & 0 \cr 0 & e^{i\delta/2}} ,
\end{equation}
with $\delta = \delta_{1} - \delta_{2}$ .
In measurement processes, the overall phase factor
$e^{i(\delta_{1} + \delta_{2})/2}$
cannot be detected, and can therefore be deleted.  The polarization
effect of the filter is solely determined by the matrix
\begin{equation}\label{shif11}
Z(\delta) = \pmatrix{e^{i\delta/2} & 0 \cr 0 & e^{-i\delta/2}} ,
\end{equation}
which leads to a phase difference of $\delta$ between the $x$ and $y$
components.  The form of this matrix is given in Eq.(\ref{herm11}), which
serves as the rotation around the $z$ axis in the Minkowski space and
time.
\par
Also along the $x$ and $y$ directions, the attenuation coefficients
could be different.  This will lead to the matrix~\cite{hkn97josa}
\begin{equation}\label{atten}
\pmatrix{e^{-\mu_{1}} & 0 \cr 0 & e^{-\mu_{2}}}
   = e^{-(\mu_{1} + \mu_{2})/2} \pmatrix{e^{\mu/2} & 0 \cr 0 & e^{-\mu/2}}
\end{equation}
with $\mu = \mu_{2} - \mu_{1}$ .
If $\mu_1 = 0$ and $\mu_2 = \infty$, the above matrix becomes
\begin{equation}\label{polar}
\pmatrix{1 & 0 \cr 0 & 0} ,
\end{equation}
which eliminates the $y$ component.  This matrix is known as a polarizer
in the textbooks~\cite{saleh07}, and is a special case of the attenuation
matrix of Eq.(\ref{atten}).

\par
This attenuation matrix tells us that the electric fields are attenuated
at two different rates.  The exponential factor $e^{-(\mu_{1} + \mu_{2})/2}$
reduces both components at the same rate and does not affect the state of
polarization.  The effect of polarization is solely determined by the
squeeze matrix~\cite{hkn97josa}
\begin{equation}\label{sq11}
B(\mu) = \pmatrix{e^{\mu/2} & 0 \cr 0 & e^{-\mu/2}} .
\end{equation}
This diagonal matrix is given in Eq.(\ref{symm11}).  In the language of
space-time symmetries, this matrix performs a Lorentz boost along the
$z$ direction.

\par
The polarization axes are not always the $x$ and $y$ axes.
For this reason, we need the rotation matrix
\begin{equation}\label{rot11}
R(\theta) = \pmatrix{\cos(\theta/2) & -\sin(\theta/2)
\cr \sin(\theta/2) & \cos(\theta/2)} ,
\end{equation}
which, according to Eq.(\ref{herm11}), corresponds to the rotation around
the $y$ axis in the space-time symmetry.
\par
Among the rotation angles, the angle of $45^o$ plays an important role in
polarization optics.  Indeed, if we rotate the squeeze matrix of Eq.(\ref{sq11})
by $45^o$, we end up with the squeeze matrix
\begin{equation}\label{sq22}
R(\theta) = \pmatrix{\cosh(\lambda/2) & \sinh(\lambda/2)
\cr \sinh(\lambda/2) & \cosh(\lambda/2)} ,
\end{equation}
which is also given in Eq.(\ref{symm11}).  In the language of space-time physics,
this matrix lead to a Lorentz boost along the $x$ axis.
\par
Indeed, the $G$ matrix of Eq.(\ref{g22}) is the most general form of the
transformation matrix applicable to the Jones matrix.  Each of the above four
matrices plays its important role in special relativity, as we discussed in
Sec.~\ref{pew}.  Their respective roles in optics and particle physics are
given in Table~\ref{tab22}.

\par

However, the Jones matrix alone cannot tell whether the two components are
coherent with each other.  In order to address this important degree of freedom,
we use the coherency matrix~\cite{azzam77,born80}
\begin{equation}\label{cocy11}
C = \pmatrix{S_{11} & S_{12} \cr S_{21} & S_{22}},
\end{equation}
with
\begin{equation}
<\psi_{i}^* \psi_{j}> = \frac{1}{T} \int_{0}^{T}\psi_{i}^* (t + \tau) \psi_{j}(t) dt,
\end{equation}
where $T$ is for a sufficiently long time interval, is much larger than $\tau$.
Then, those four elements become~\cite{hkn97}
\begin{eqnarray}
&{}& S_{11} = <\psi_{1}^{*}\psi_{1}> =a^2  , \qquad
S_{12} = <\psi_{1}^{*}\psi_{2}> = ab~e^{-(\sigma +i\delta)} , \nonumber \\[1ex]
&{}& S_{21} = <\psi_{2}^{*}\psi_{1}> = ab~e^{-(\sigma -i\delta)} ,  \qquad
S_{22} = <\psi_{2}^{*}\psi_{2}>  = b^2 .
\end{eqnarray}
The diagonal elements are the absolute values of $\psi_1$ and $\psi_2$
respectively.  The off-diagonal elements could be smaller than the product
of $\psi_1$ and $\psi_2$, if the two beams are not completely coherent.
The $\sigma$ parameter specifies the degree of coherency.
\par
This coherency matrix is not always real but it is Hermitian.  Thus it can be
diagonalized by a unitary transformation.  If this matrix is normalized so
that its trace is one, it becomes a density matrix~\cite{fey72}.
\par

\begin{table}
\caption{Polarization optics and special relativity sharing the same mathematics.
Each matrix has its clear role in both optics and relativity.  The determinant
of the Stokes or the four-momentum matrix remains invariant under Lorentz
transformations.  It is interesting to note that the decoherence parameter
(least fundamental) in optics corresponds to the mass (most fundamental) in
particle physics.}\label{tab22}
\vspace{2mm}

\begin{center}
\begin{tabular}{lcl}
\hline
\hline \\[0.5ex]
 Polarization Optics & Transformation Matrix  &  Particle Symmetry \\[1.0ex]
\hline \\
Phase shift $\delta$  &
$\pmatrix{e^{\delta/2} & 0\cr 0 & e^{-i\delta/2}}$
&  Rotation around $z$.
\\[4ex]
Rotation around $z$  &
$\pmatrix{\cos(\theta/2) & -\sin(\theta/2)\cr \sin(\theta/2) & \cos(\theta/2)}$
&  Rotation around  $y$.
\\[4ex]
Squeeze along $x$ and $y$  &
$\pmatrix{e^{\eta/2} & 0\cr 0 & e^{-\eta/2}}$
&  Boost along $z$.
\\[4ex]
Squeeze along $45^o$  &
$\pmatrix{\cosh(\lambda/2) & \sinh(\lambda/2)\cr \sinh(\lambda/2)
                & \cosh(\lambda/2)}$
&   Boost along $x$.
\\[4ex]
$(ab)^{2} \sin^2\chi$  & Determinant &  (mass)$^2$
\\[4ex]
\hline
\hline\\[-0.8ex]
\end{tabular}
\end{center}
\end{table}

If we start with the Jones vector of the form of Eq.(\ref{jvec11}), the coherency
matrix becomes
\begin{equation}\label{cocy22}
C = \pmatrix{a^2 & ab~e^{-(\sigma + i\delta)} \cr
ab~e^{-(\sigma - i\delta)} & b^2} .
\end{equation}
We are interested in the symmetry properties of this matrix.  Since the
transformation matrix applicable to the Jones vector is the two-by-two
representation of the Lorentz group, we are particularly interested in the
transformation matrices applicable to this coherency matrix.
\par
The trace and the determinant of the above coherency matrix
are
\begin{eqnarray}
&{}& \det(C) = (ab)^2 \left(1 - e^{-2\sigma}\right), \nonumber \\[2ex]
&{}& \mbox{tr}(C) = a^2 + b^2 .
\end{eqnarray}
Since $e^{-\sigma}$ is always smaller than one, we can introduce
an angle $\chi$ defined as
\begin{equation}
\cos\chi = e^{-\sigma} ,
\end{equation}
and call it the ``decoherence angle.''  If $\chi = 0$, the decoherence is
minimum, and it is maximum when $\chi = 90^o$.  We can then write the
coherency matrix of Eq.(\ref{cocy22}) as
\begin{equation}\label{cocy22b}
C = \pmatrix{a^2 & ab(\cos\chi)e^{-i\delta} \cr
ab(\cos\chi)e^{i\delta} & b^2}.
\end{equation}

\par
The degree of polarization is defined as~\cite{saleh07}
\begin{equation}\label{pdegree}
f = \sqrt{ 1 - \frac{4~\det(C)}{(\mbox{tr}(C))^2}} =
        \sqrt{1 - \frac{4(ab)^2\sin^2\chi}{(a^2 + b^2)^2}} .
\end{equation}
This degree is one if $\chi = 0$.  When $\chi = 90^o$, it becomes
\begin{equation}
  \frac{a^2 - b^2}{a^2 + b^2} ,
\end{equation}
Without loss of generality, we can assume that $a$ is greater than $b$.
If they are equal, this minimum degree of polarization is zero.
\par

Under the influence of the Naimark transformation given in Eq.(\ref{naim}),
this coherency matrix is transformed as
\begin{eqnarray}\label{trans22}
&{}& C' = G~C~G^{\dagger} =
\pmatrix{S'_{11} & S'_{12} \cr S'_{21} & S'_{22}} \nonumber \\[2ex]
&{}&\hspace{5ex} = \pmatrix{\alpha & \beta \cr \gamma & \delta}
\pmatrix{S_{11} & S_{12} \cr S_{21} & S_{22}}
\pmatrix{\alpha^{*} & \gamma^{*} \cr \beta^{*} & \delta^{*}} .
\end{eqnarray}

\par
It is more convenient to make the following linear combinations.
\begin{eqnarray}\label{stokes11}
&{}& S_{0} = \frac{S_{11} + S_{22}}{\sqrt{2}},  \qquad
    S_{3} = \frac{S_{11} - S_{22}}{\sqrt{2}},    \nonumber \\[2ex]
&{}& S_{1} = \frac{S_{12} + S_{21}}{\sqrt{2}}, \qquad
S_{2} = \frac{S_{12} - S_{21}}{\sqrt{2} i}.
\end{eqnarray}
These four parameters are called Stokes parameters, and four-by-four
transformations applicable to these parameters are widely known as
Mueller matrices~\cite{azzam77,bros98}.
However, if the Naimark transformation given in Eq.(\ref{trans22}) is
translated into the four-by-four Lorentz transformations according to the
correspondence given in the Appendix,  the Mueller matrices constitute
a representation of the Lorentz group.

\par

Another interesting aspect of the two-by-two matrix formalism is that
the coherency matrix can be formulated in terms of the
quarternions~\cite{dlugu09,tudor10}.  The quarternion representation
can be translated into rotations in the four-dimensional space.
There is a long history between the Lorentz group and the four-dimensional
rotation group.  It would be interesting to see what the quarternion
representation of polarization optics will add to this history between
those two similar but different groups.
\par

As for earlier applications of the two-by-two representation of the Lorentz
group, we note the vector representation by Fedorov~\cite{fedo70,fedo79}.
Fedorov showed that it is easier to carry our kinematical calculations
using his two-by-two representation.  For instance, the computation of
the Wigner rotation angle is possible in the two-by-two
representation~\cite{bk05jpa}.

\section{Geometry of the Poincar\'e Sphere}\label{poincs}

We now have the four-vector $\left(S_0, S_3, S_1, S_2\right)$, which is
Lorentz-transformed like the space-time four-vector $(t, z, x, y)$ or the
energy-momentum four vector of Eq.(\ref{momen11}).  This Stokes four-vector
has a three-component subspace $\left(S_3, S_1, S_2\right)$, which is like
the three-dimensional Euclidean subspace in the four-dimensional Minkowski
space.  In this three-dimensional subspace, we can introduce the spherical
coordinate system with
\begin{eqnarray}
&{}&  R = \sqrt{S_3^2 + S_1^2 + S_2^2} \nonumber\\[1ex]
&{}&  S_3 = R\cos\xi,       \nonumber \\[1ex]
&{}&  S_1 = R ( \sin\xi) \cos\delta, \qquad S_2 = R (\sin\xi) \sin\delta .
\end{eqnarray}
\par

\begin{figure}
\centerline{\includegraphics[scale=0.37]{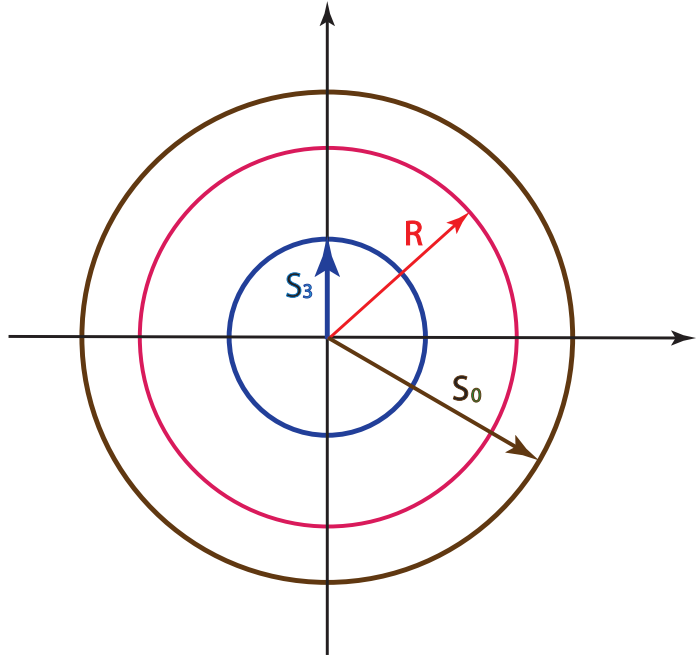}}
\caption{Radius of  the Poincar\'e sphere.  The radius $R$ takes its maximum
value $S_0$ when the decoherence angle $\chi$ is zero.  It takes its
minimum value $S_z$ when $\chi = 90^o.$  The degree of polarization is
maximum when
$R = S_0$, and is minimum when $R = S_z$. }\label{poincs22}
\end{figure}
\par
The radius $R$ is the radius of the traditional Poincar\'e sphere, and is
\begin{equation}\label{radius11}
 R = \frac{1}{2} \sqrt{(a^2 - b^2)^2 + 4(ab)^2\cos^{2}\chi} .
\end{equation}
with
\begin{equation}
S_3 = \frac{a^2 - b^2}{2}.
\end{equation}
This spherical picture of is traditionally known as the
Poincar\'e sphere~\cite{azzam77,born80,bros98}
\par
The radius $R$ takes its maximum value $S_0$ when $\chi = 0^o$.  It
decreases and reaches its minimum value, $S_3$, when $\chi = 90^o$.
In terms of $R$, the degree of polarization given in Eq.(\ref{pdegree})
is
\begin{equation}
f = \frac{R}{S_0} .
\end{equation}
This aspect of the radius R is illustrated in Fig.~\ref{poincs22}.

\par

Under the Lorentz transformation, all four Stokes parameters are subject
to change.  The maximum radius $S_0$ and the minimum radius $S_3$ do not
remain invariant.  The radius $R$ is also subject to change.  These three
radii are illustrated in Fig.~\ref{poincs22}.  While Lorentz transformations
shake up these parameters, is there any quantity remaining invariant?
\par
Let us go back to the four-momentum matrix of Eq.(\ref{momen11}).  Its
determinant is $m^2$ and remains invariant.   Likewise, the determinant
of the coherency matrix of Eq.\ref{cocy22b} should also remain invariant.
The determinant in this case is
\begin{equation}
 S_0^2 - R^2 = (ab)^2 \sin^2\chi .
\end{equation}
This quantity remains invariant.  This aspect is shown on the last row
of Table~\ref{tab22}.
\par
While the decoherence parameter is not fundamental is influenced by
environment, it plays the same mathematical role as in the particle mass
which remains as the most fundamental quantity since Isaac Newton,
and even after Einstein.

\section{Entropy and Feynman's Rest of the Universe}\label{restof}

Another way to measure the lack of coherence is to compute the entropy
of the system.  The coherency matrix of Eq.(\ref{cocy22}) leads to the
density matrix of the form
\begin{equation}\label{den11}
 \rho(\chi) = \frac{1}{a^2 + b^2}   \pmatrix{a^2 & (ab)e^{-i\delta}(\cos\chi)
\cr  (ab)e^{i\delta}(\cos\chi) &  b^2} ,
\end{equation}
whose trace is one.  This matrix can be diagonalized to
\begin{equation}\label{den22}
 \rho(\chi) = \frac{1}{2} \pmatrix{1 + f  & 0  \cr  0 &  1 - f} ,
\end{equation}
where $f$ is the degree of polarization given in Eq.(\ref{pdegree}), which
can also be written as
\begin{equation}
f = \frac{\sqrt{\left(a^2 + b^2\right)^2 - 4(ab)^2\sin^2\chi}} {a^2 + b^2 }.
\end{equation}
As we noted before,  this quantity is one when $\chi = 0$ and takes its minimum
value when $\chi = 90^o$.

\par
Then, the entropy becomes
\begin{equation}
S = -\left(\frac{1 + f}{2} \right) \ln{\left(\frac{1 + f}{2}\right)}
 - \left(\frac{1 - f}{2} \right) \ln{\left(\frac{1 - f}{2}\right)}.
\end{equation}
The entropy $S$ becomes zero when $\chi = 0$.  When $\chi = 90^o$, the entropy
takes its maximum value
\begin{equation}
 \frac{a^2}{a^2 + b^2} \ln{\left(\frac{a^2 + b^2}{a^2}\right)}
 + \frac{b^2}{a^2 + b^2} \ln{\left(\frac{a^2 + b^2}{b^2}\right)},
\end{equation}
\par
In the special case of $a = b$,
\begin{equation}
  S = -\left[\cos^2\left(\frac{\chi}{2}\right)\right]
         \ln\left[\cos^2\left(\frac{\chi}{2}\right)\right]
    -\left[\sin^2\left(\frac{\chi}{2}\right)\right]
         \ln\left[\sin^2\left(\frac{\chi}{2}\right)\right].
\end{equation}
This entropy is a monotonically increasing function.  It is zero when $\chi = 0$.  \
Its maximum value is $2 \ln(2)$ when $\chi = 90^o$.

\par

The symmetry between $\cos\chi$ and $\sin\chi$ is well known.  Let us now
consider another Poincar\'e sphere where $\cos\chi$ is replaced by $\sin\chi$.
Then the density matrix and entropy become
\begin{eqnarray}\label{den33}
&{}& \rho'(\chi) = \frac{1}{2} \pmatrix{1 + f'  & 0  \cr  0 &  1 - f'} ,    \nonumber \\[1.0ex]
&{}&  S' = -\left(\frac{1 + f'}{2} \right) \ln{\left(\frac{1 + f'}{2}\right)}
 - \left(\frac{1 - f'}{2} \right) \ln{\left(\frac{1 - f'}{2}\right)},
\end{eqnarray}
respectively, with
\begin{equation}
f' = \frac{\sqrt{\left(a^2 - b^2\right)^2 + 4(ab)^2\sin^2\chi}} {a^2 + b^2 } ,
\end{equation}
which takes its minimum and maximum values when $\chi = 0$ and $\chi = 90^o$
respectively.
\par
Indeed, $S$ and $S'$ move in opposite directions as $\chi$ changes.  Thus we are
led to consider their addition $\left(S + S'\right)$.  The question is whether
it remains constant.  The answer is No.
On the other hand, the determinant of the first coherency matrix is
$(ab)^2\sin^2\chi,$ and it is $(ab)^2\cos^\chi$ for the second determinant.
Thus the addition of these two determinants
\begin{equation}\label{add}
(ab)^2\sin^2\chi + (ab)^2\cos^\chi = (ab)^2 .
\end{equation}
is independent of the decoherence angle $\chi$.

\par
While the determinant of is a Lorentz-invariant quantity, there could be a
larger group which will change the value of the determinant, thus the decoherence
angle.  This question has been addressed in the literature~\cite{bk06jpa}.
\par
Then there comes the issue of Feynman's rest of the universe.  In his book on
statistical mechanics~\cite{fey72}, Feynman divides the quantum universe into
two systems, namely the world in which we do physics, and the rest of the
universe beyond our scope.
\par
However, we could gain a better understanding of physics if we can construct
a model of the rest of the universe which can be explained in terms of the
physical laws applicable to the world which we observe.  With this point in
mind, Han {\it et al.} considered a system of two coupled world~\cite{hkn99ajp}.
One of those oscillators belong to the world in which we do physics, while
the other is in the rest of the universe.  This constitute of an illustrative
example of Feynman's rest of the universe.
\par
In this section, we discussed two Poincar\'e spheres which are coupled by
the addition formula of Eq.(\ref{add}).  One of the spheres talks about
the physical world, and the other takes care of the entropy variation of
this world through the conservation of Eq.(\ref{add}).  Indeed, these two
coupled Poincar\'e spheres constitute another illustrative example of
Feynman's rest of the universe.

\section*{Concluding Remarks}
In this report, we noted first that the group of Lorentz transformations
can be formulated in terms of two-by-two matrices.  This two-by-two formalism
can also be used for transformations of the coherency matrix in polarization
optics.
\par
Thus, this set of four Stokes parameters is like a Minkowskian four-vector
under four-by-four Lorentz transformations.  The geometry of the Poincar\'e
sphere can be extended to accommodate this four-dimensional transformations.
\par
It is shown that the decoherence parameter in the Stokes formalism is invariant
under Lorentz transformations, like the particle mass in Einstein's four-vector
formalism of the energy and momentum.

\section*{Acknowledgments}
First of all, I would like to thank Professor Sergei Kilin for inviting me
to the International Conference ``Spins and Photonic Beams at Interface,''
honoring Academician F. I. Fedorov.  In addition to numerous original
contributions in optics, Fedorov wrote a book on two-by-two representations
of the Lorentz group based on his own research in this subject.  It was
quite appropriate for me to present a paper on applications of the Lorentz
group to optical science.  I would like thank Professors V. A. Dluganovich
and M. Glaynskii for bringing to my attention the papers and the book
written by Academician Fedorov, as well as their own papers.

\begin{appendix}

\section*{Appendix}

In Sec.~\ref{pew}, we listed four two-by-two matrices whose repeated applications
lead to the most general form of the two-by-two matrix $G$.  It is known  that
every $G$ matrix can be translated into a four-by-four Lorentz transformation
matrix through~\cite{naimark54,knp86,hkn97}
\begin{equation}\label{trans44a}
\pmatrix{t' + z' \cr x' - iy'  \cr x' + iy' \cr t' - z'} =
\pmatrix{\alpha\alpha^{*} & \alpha \beta^{*} &
\beta\alpha^{*} & \beta \beta^{*} \cr
\alpha \gamma^{*} & \alpha \delta^{*} &
\beta \gamma^{*} & \beta \delta^{*} \cr
\gamma \alpha^{*}  & \gamma \beta^{*} &
\delta \alpha^{*} & \delta \beta^{*} \cr
\gamma \gamma^{*} & \gamma \delta^{*} &
\delta \gamma^{*} & \delta \delta^{*}}
\pmatrix{t + z \cr x - iy  \cr x + iy \cr t - z} ,
\end{equation}
\par

and
\begin{equation}\label{trans44b}
\pmatrix{t \cr z  \cr x \cr y } =  \frac{1}{2}
\pmatrix{1 & 0 & 0 & 1 \cr 1 & 0 & 0 & -1 \cr 0  & 1 & 1 & 0 \cr
0 & i & -i & 0}
\pmatrix{t + z \cr x - iy  \cr x + iy \cr t - z} .
\end{equation}

\par
These matrices appear to be complicated, but it is enough to study the
our  matrices of Eq.(\ref{herm11})  and Eq.(\ref{symm11}) enough
to cover all the matrices in this group.  Thus, we give their four-by-four
equivalents in this appendix.
\begin{equation}
Z(\delta) = \pmatrix{e^{i\delta/2} & 0 \cr 0 & e^{-i\delta/2}}
\end{equation}
leads to the four-by-four matrix
\begin{equation}
\pmatrix{1 & 0 & 0 & 0 \cr 1 & 0 & 0 & 0 \cr 0  & 1 & \cos \delta &  -\sin\delta \cr
    0 & 0 & \sin\delta & \cos\delta}.
\end{equation}

Likewise,
\begin{equation}
B(\mu) = \pmatrix{e^{\mu/2} & 0 \cr 0 & e^{-\mu/2} }
 \rightarrow
\pmatrix{\cosh\mu &  \sinh\mu & 0 & 0 \cr
    \sinh\mu & \cosh\mu & 0 &  0 \cr
   0  & 0 & 1 & 0 \cr   0 & 0 & 0 & 1},
\end{equation}

\begin{equation}
R(\theta) = \pmatrix{\cos(\theta/2) & -\sin(\theta/2) \cr \sin(\theta/2) & \sin(\theta/2) }
\rightarrow
\pmatrix{1 & 0 & 0 & 0 \cr
    0 & \cos\theta & -\sin\theta &  0 \cr
   0  & \sin\theta & \cos\theta & 0 \cr   0 & 0 & 0 & 1},
\end{equation}
and

\begin{equation}
S(\lambda) = \pmatrix{\cosh(\lambda/2) & \sinh(\lambda/2) \cr \sinh(\lambda/2) & \sinh(\lambda/2) }
\rightarrow
\pmatrix{\cosh\lambda & 0 & \sinh\lambda & 0 \cr
    0 & 1 & 0 &  0 \cr
   \sinh\lambda  & 0 & \cosh\lambda & 0 \cr
     0 & 0 & 0 & 1}.
\end{equation}
\end{appendix}

\end{document}